\documentclass[12pt]{article}
\usepackage{graphicx}

\def\pbnr{UCHEP-13-04}
\def\today{December 22, 2013}
\def\speaker{A. J. Schwartz}
\def\title{\cp\ Asymmetry Measurements in $D$ Decays from Belle}
\def\affiliation{Physics Departement\\
University of Cincinnati, P.O. Box 210011, Cincinnati, Ohio 45221 USA}

\newcommand\pubnumber{\pbnr}
\newcommand\pubdate{\today}
\def\Title#1{\begin{center} {\Large #1 } \end{center}}
\def\Author#1{\begin{center}{ \sc #1} \end{center}}

\newcommand{\OnBehalf}[1]{\sbox0{#1}\ifdim\wd0=0pt
        {}
	\else
	{\\on behalf of #1}
	\fi}
\newcommand{\SupportedBy}[1]{\sbox0{#1}\ifdim\wd0=0pt
        {}
	\else
	{\footnote{#1}}
	\fi}
\def\Address#1{\begin{center}{ \it #1} \end{center}}

\newcommand\pubblock{\includegraphics[width=5cm]{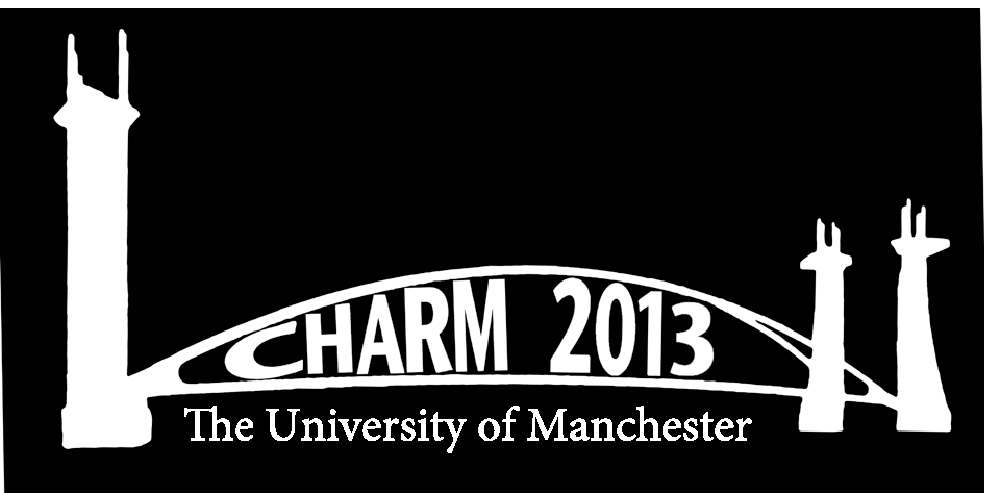}\hfill{\begin{tabular}{l} \pubnumber\\
         \pubdate  \end{tabular}}}
\newenvironment{Abstract}{\begin{quotation}  }{\end{quotation}}
\newenvironment{Presented}{\begin{quotation} \begin{center} 
             PRESENTED AT\end{center}\bigskip 
      \begin{center}\begin{large}}{\end{large}\end{center} \end{quotation}}
\def\Acknowledgements{\bigskip  \bigskip \begin{center} \begin{large}
             \bf ACKNOWLEDGEMENTS \end{large}\end{center}}
\def\venue{The 6$^{th}$ International Workshop on Charm Physics\\
(CHARM 2013)\\
Manchester, UK,  31 August -- 4 September, 2013}

\def\beq{\begin{equation}}
\def\eeq#1{\label{#1}\end{equation}}
\def\eeqn{\end{equation}}

\def\beqa{\begin{eqnarray}}
\def\eeqa#1{\label{#1}\end{eqnarray}}
\def\eeqan{\end{eqnarray}}

\let\bar=\overbar

\def\Dslash{\not{\hbox{\kern-4pt $D$}}}
\def\dslash{\not{\hbox{\kern-2pt $\del$}}}

\def\msb{{\bar{\ssstyle M \kern -1pt S}}}

\def\cp{$CP$\/}
\def\cpv{$CPV$\/}

\def\ycp{$y^{}_{CP}$\/}
\def\agamma{$A^{}_{\Gamma}$\/}

\def\ra{\!\rightarrow\!}
\def\kbar{\overline{K}{}^{\,0}}
\def\dbar{\overline{D}{}^{\,0}}
\def\bbar{\overline{B}{}^{\,0}}

\def\acpKK{A^{KK}_{CP}}
\def\acpPP{A^{\pi\pi}_{CP}}
\def\acpKz{A^{\kbar}_{CP}}
\def\acpKzbKp{A^{\kbar K^+}_{CP}}
\def\acpKzbPp{A^{\kbar\pi^+}_{CP}}

\begin{document}
\begin{titlepage}
\pubblock

\vfill
\Title{\title}
\vfill
\Author{\speaker}
\Address{\affiliation}
\vfill
\begin{Abstract}
We present measurements of \cp\ asymmetries in 
$D$ decays performed by the Belle experiment running
at the KEKB asymmetric-energy $e^+e^-$ collider.
\end{Abstract}
\vfill
\begin{Presented}
\venue
\end{Presented}
\vfill
\end{titlepage}
\def\thefootnote{\fnsymbol{footnote}}
\setcounter{footnote}{0}
\setcounter{page}{2}

\section{Introduction}

The phenomenon of \cp\ violation (\cpv) is well-established 
in the $K^0$-$\kbar$ and  $B^0$-$\bbar$ systems~\cite{cpv_kk,cpv_bb}. 
The rate observed confirms the Cabibbo-Kobayashi-Maskawa 
(CKM) theory of quark flavor mixing~\cite{ckm}. The CKM theory 
predicts only tiny \cpv\ in the $D^0$-$\dbar$ system~\cite{GrossmanKaganNir}, 
and to-date such \cpv\ has not been observed. Both time-independent 
and time-dependent measurements of partial widths can exhibit 
\cp\ asymmetries. The former results mainly from interference 
between two decay amplitudes with different weak phases; this 
is called ``direct'' \cpv. The latter results from either unequal 
rates of flavor mixing (called ``indirect'' \cpv) or interference 
between a mixed and an unmixed decay amplitude. In all cases new 
physics can increase the rate of \cpv\ significantly above that 
predicted by the CKM theory~\cite{GrossmanKaganNir}. 
Here we present results from searches for \cpv\ in $D$ decays from 
the Belle experiment. For a review of mixing and \cpv\ formalism, 
see Ref.~\cite{KirkbyNir}.

\section{\boldmath Time-dependent $D^0(t)\ra K^+K^-/\pi^+\pi^-$}

\begin{figure}[htb]
\centering
\includegraphics[width=5.5in]{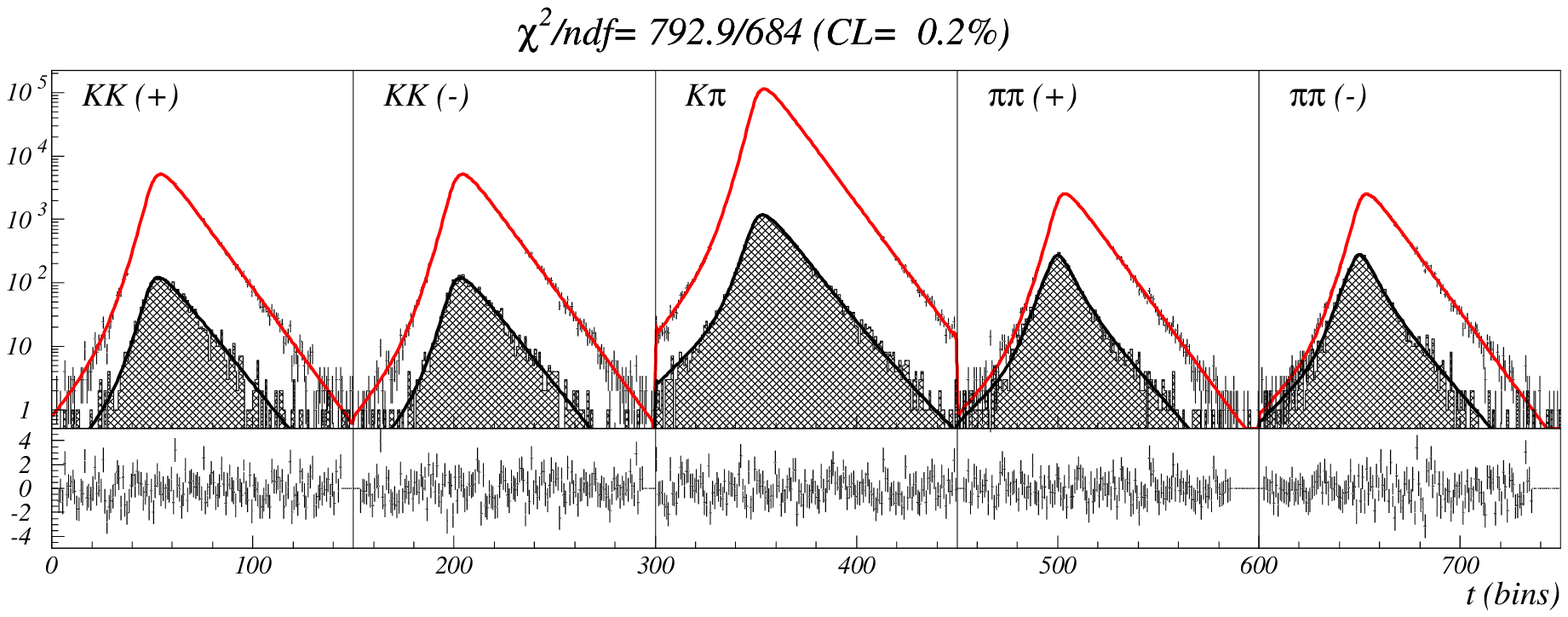}
\vskip-0.25in
\caption{Decay time distributions for $D^0\ra K^+K^-/\pi^+\pi^-$,
$\dbar\ra K^+K^-/\pi^+\pi^-$, and $D^0\ra K^-\pi^+$. Fitting these 
distributions yields $y^{}_{CP}$ and $A^{}_\Gamma$ via 
Eqs.~(\ref{eqn:ycp_exp}) and (\ref{eqn:agamma_exp}).
The shaded histograms show background contributions; 
the lower plots show the fit residuals.}
\label{fig:ycp_staric1}
\end{figure}

\begin{figure}[htb]
\centering
\includegraphics[width=5.5in]{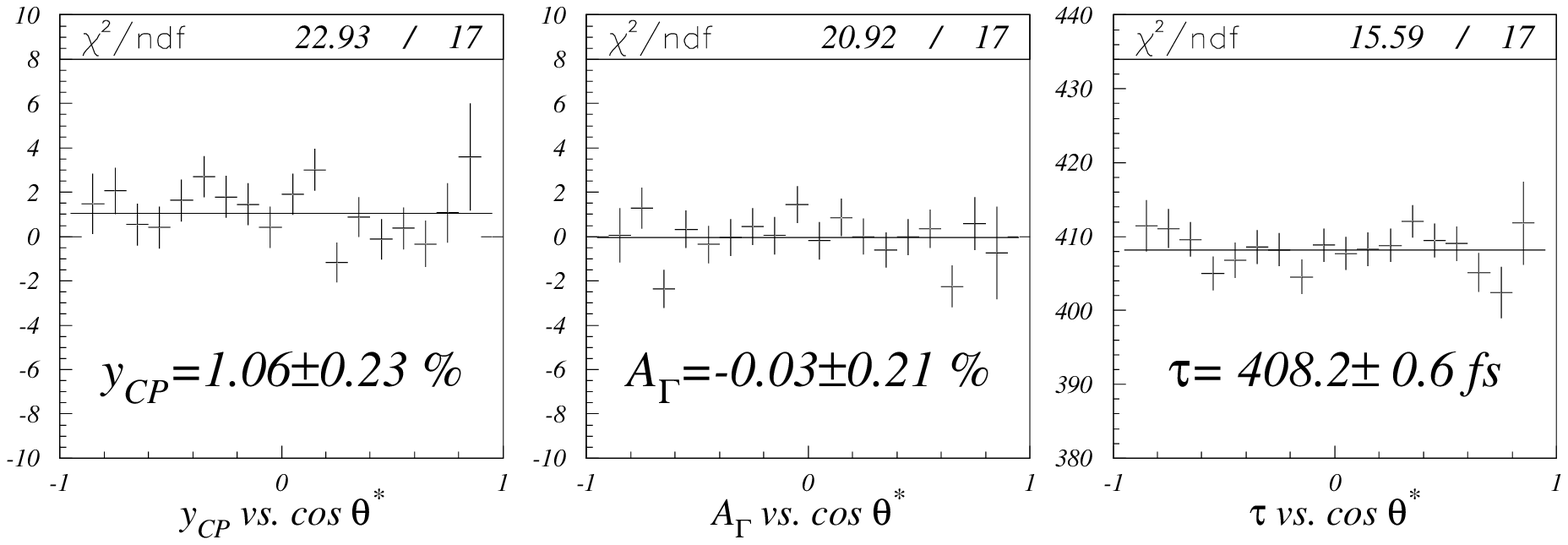}
\vskip-0.20in
\caption{Fitted values of $y^{}_{CP}$, $A^{}_\Gamma$, and
$\tau^{}_{D^0}$ for SVD2 data in 18 bins of $\cos\theta^*$ 
(see text). Fitting these points to constants yields the 
overall values listed.}
\label{fig:ycp_staric2}
\end{figure}

The Belle experiment has measured the mixing parameter 
\ycp\ and the \cp-violating parameter \agamma\ using 
977~fb$^{-1}$ of data~\cite{belle_staric}.
Both observables depend on mixing parameters 
$x$ and $y$ and \cpv\ parameters $|q/p|$ and 
$\phi$. To lowest order the relations are
\begin{eqnarray}
y^{}_{CP} & = & 
\frac{(|q/p|+|p/q|)}{2}y\cos\phi \ -\ \frac{(|q/p|-|p/q|)}{2}x\sin\phi 
\label{eqn:ycp} \\
A^{}_\Gamma & = & 
\frac{(|q/p|-|p/q|)}{2}y\cos\phi \ -\ \frac{(|q/p|+|p/q|)}{2}x\sin\phi 
\label{eqn:agamma}
\end{eqnarray}
The parameters are determined by measuring lifetimes
of $D^0$ and $\dbar$ mesons to flavor-specific and \cp-specific final
states, e.g.,
\begin{eqnarray}
y^{}_{CP} & = & \frac{\tau(K^-\pi^+)}{\tau(K^+ K^-)}-1 
\label{eqn:ycp_exp} \\
A^{}_\Gamma & = & \frac{\tau(\dbar\ra K^+ K^-)-\tau(D^0\ra K^+ K^-)}
{\tau(\dbar\ra K^+ K^-)+\tau(D^0\ra K^+ K^-)}\,.
\label{eqn:agamma_exp}
\end{eqnarray}
The latter measurement requires tagging the flavor of the decaying
$D$ meson, and this is done by reconstructing $D^{*+}\ra D^0\pi^+$ and
$D^{*-}\ra\dbar\pi^-$ decays, i.e., the charge of the accompanying
$\pi^\pm$ (which has low momentum and is often called the
``slow pion'') identifies the $D$ flavor.
Both $K^+K^-$ and $\pi^+\pi^-$ final states are used by Belle,
and the fitted lifetime distributions are shown in Fig.~\ref{fig:ycp_staric1}.
The precision of the measurement depends upon good understanding of
the decay time resolution of the detector. For Belle, the resolution 
function depends upon $\theta^*$, the polar angle with respect to 
the $e^+$ beam of the $D^0$ in the $e^+e^-$ center-of-mass (CM) frame. 
Thus the ratios in Eqs.~(\ref{eqn:ycp_exp}) and (\ref{eqn:agamma_exp}) 
are measured in bins of $\cos\theta^*$. The resolution function also
depends on the detector configuration: 
for the first 153~fb$^{-1}$ of data a 3-layer silicon vertex detector (SVD) 
was used, while for the remaining data a 4-layer SVD was used. These 
running periods (``SVD1'' and ``SVD2'') are treated separately.
The results for SVD2 data are plotted in Fig.~\ref{fig:ycp_staric2}. 
Fitting both SVD1 and SVD2 values to constants gives
\begin{eqnarray}
y^{}_{CP} & = & (+1.11\pm 0.22\pm 0.11)\% \\
A^{}_\Gamma & = & (-0.03\pm 0.20\pm 0.08)\% \,.
\end{eqnarray}
As a test of the resolution function, the absolute lifetime
for $D^0\ra K^-\pi^+$~\cite{charge-conjugates} decays is also 
measured. The result is $408.5\pm 0.5$~fs, which is consistent 
with the world average~\cite{PDG}.

\section{\boldmath Time-integrated $D^0\ra K^+K^-/\pi^+\pi^-$}
\label{sec:staric_ti}

The $D\ra K^+K^-/\pi^+\pi^-$ samples used previously can be 
integrated over all decay times to measure the \cp\ asymmetry 
$A^f_{CP}$, defined as
\begin{eqnarray}
A^f_{CP} & \equiv & 
\frac{\Gamma(D^0\ra f) - \Gamma(\dbar\ra f)}
{\Gamma(D^0\ra f) + \Gamma(\dbar\ra f)}.
\label{eqn:cpasymm}
\end{eqnarray}
This parameter is a difference in partial widths rather 
than a difference in lifetimes and thus depends strongly on the 
specific final state. In addition to the underlying \cp\ asymmetry, 
there is a ``forward-backward'' asymmetry ($A^{}_{FB}$) in 
$D^0/\dbar$ production due to $\gamma$-$Z^0$ electroweak
interference and higher order QED effects in $e^+e^-\ra c\bar{c}$;
and there is an asymmetry $A^\pi_\varepsilon$ 
in the reconstruction of $\pi^{\pm}_s$ from $D^{*\pm}\ra D\pi^\pm_s$ 
decays used to tag the $D$ flavor. The reconstructed asymmetry one 
measures is the sum of all three:
$A^f_{\rm recon} = A^f_{CP} + A^{}_{FB} + A^\pi_\varepsilon$.
Belle has measured $A^f_{\rm recon}$ for $D^0\ra K^+K^-/\pi^+\pi^-$
decays using 977~fb$^{-1}$ of data~\cite{belle_staric_acp} to 
determine $\acpKK$ and $\acpPP$.

\begin{figure}[htb]
\centering
\includegraphics[width=4.5in]{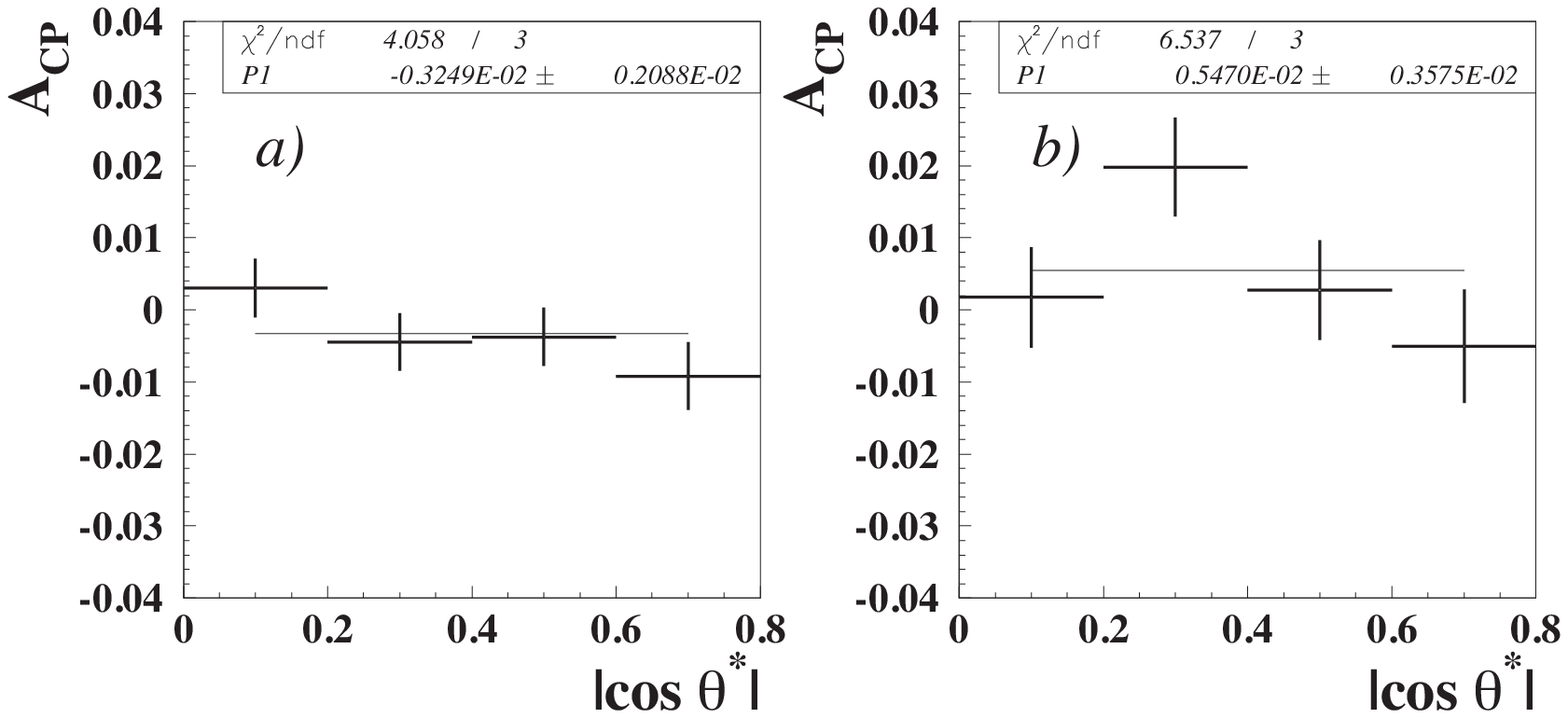}
\vskip-0.10in
\caption{Asymmetries $\acpKK$ (left) and $\acpPP$ (right)
calculated in bins of $\cos\theta^*$ (see text). Fitting these
values to constants yields the results listed in 
Eqs.~(\ref{eqn:akk}) and (\ref{eqn:apipi}).}
\label{fig:Acp_ko}
\end{figure}

To correct for $A^\pi_\varepsilon$, Belle measures $A^f_{\rm recon}$ 
for flavor-tagged and untagged $D^0\ra K^-\pi^+$ decays. These 
decays have an additional asymmetry due to differences in the 
reconstruction efficiency of $K^-\pi^+$ versus $K^+\pi^-$;
this difference is denoted $A^{K\pi}_\varepsilon$. Thus
\begin{eqnarray}
A^{K\pi}_{\rm tagged} & = & A^{K\pi}_{CP} + A^{}_{FB} + 
A^{K\pi}_\varepsilon + A^\pi_\varepsilon \\
A^{K\pi}_{\rm untagged} & = & A^{K\pi}_{CP} + A^{}_{FB} + 
A^{K\pi}_\varepsilon\,,
\end{eqnarray}
and taking the difference 
$A^{K\pi}_{\rm tagged} -A^{K\pi}_{\rm untagged}$ yields $A^\pi_\varepsilon$.
In practice this is done by re-weighting events: 
$A^{K\pi}_{\rm tagged}$ is calculated by weighting 
$D^0$ decays by a factor
$1 - A^{K\pi}_{\rm untagged}(p^{}_{D^0}, \cos\theta^{}_{D^0})$
and weighting $\dbar$ decays by a factor 
$1 + A^{K\pi}_{\rm untagged}(p^{}_{\dbar}, \cos\theta^{}_{\dbar})$, 
where $p^{}_{D}$ and $\theta^{}_{D}$ are the momentum and
polar angle with respect to the $e^+$ beam of the~$D$. 
The resulting $A^{K\pi}_{\rm tagged}$ equals $A^\pi_\varepsilon$.
The signal asymmetry $A^f_{\rm recon}$ is then calculated by weighting 
$D^0$ decays by a factor $1 - A^\pi_\varepsilon(p^{}_\pi, \cos\theta^{}_\pi)$ 
and $\dbar$ decays by a factor
$1 + A^\pi_\varepsilon(p^{}_\pi, \cos\theta^{}_\pi)$.
The asymmetry $A^\pi_\varepsilon$ is calculated in bins 
of $p^{}_\pi$ and $\theta^{}_\pi$ to reduce systematic errors.
The result equals $A^f_{CP} + A^{}_{FB}$.
Since $A^{}_{FB}$ is an odd function of $\cos\theta^*$, where
$\theta^*$ is the polar angle of the $D$ in the CM frame,
and $A^f_{CP}$ is nominally an even function of $\theta^*$, 
the individual asymmetries are extracted via
\begin{eqnarray}
A^f_{CP} & = &
 \frac{A^{f,{\rm corr}}_{\rm recon}(\cos\theta^*) \ +\  
   A^{f,{\rm corr}}_{\rm recon}(-\cos\theta^*)}{2} \label{eqn:acp} \\
 & & \nonumber \\
A^{}_{FB} & = &
 \frac{A^{f,{\rm corr}}_{\rm recon}(\cos\theta^*) \ -\  
   A^{f,{\rm corr}}_{\rm recon}(-\cos\theta^*)}{2}\,. \label{eqn:afb}
\end{eqnarray}
The results of Eq.~(\ref{eqn:acp}) for $K^+K^-$ and 
$\pi^+\pi^-$ final states are plotted in Fig.~\ref{fig:Acp_ko}
for each bin of $\cos\theta^*$. Fitting these values 
to constants yields
\begin{eqnarray}
\acpKK & = & (-0.32\pm 0.21\pm 0.09)\% 
\label{eqn:akk} \\ 
\acpPP & = & (+0.55\pm 0.36\pm 0.09)\%\,,
\label{eqn:apipi}
\end{eqnarray}
and $\Delta A^{}_{CP} \equiv 
\acpKK - \acpPP\ =\ (-0.87\pm 0.41\pm 0.06)\%$.

\section{\boldmath Direct \cpv\ in Neutral $D^0$ Decays}

The above measurements of $A^{}_\Gamma$, $\acpKK$, and $\acpPP$ 
are sensitive to underlying parameters~\cite{GrossmanKaganNir}
\begin{eqnarray}
a^{\rm indir}_{CP} & = & 
\frac{(|q/p|+|p/q|)}{2}x\sin\phi\ -\ \frac{(|q/p|-|p/q|)}{2}y\cos\phi  
\label{eqn:acp_ind} \\
a^{\rm dir}_{CP} & = & 
2\left|\frac{\overline{\cal A}}{\cal A}\right|\sin\phi\,\sin(\bar{\delta}-\delta)\,,
\label{eqn:acp_dir} 
\end{eqnarray}
where 
${\cal A}\equiv{\cal A}(D^0\ra h^+h^-)$,
$\overline{\cal A}\equiv{\cal A}(\dbar\ra h^+h^-)$, 
and $\delta (\bar{\delta})$ is the strong phase for 
amplitude ${\cal A} (\overline{\cal A})$.
Parameters $a^{\rm indir}_{CP}$ and $a^{\rm dir}_{CP}$ 
parameterize the amounts of indirect and direct \cp\ violation in $D$
decays, respectively.  Since $D$ decays are well-dominated
by tree amplitudes, the phase 
$\phi\equiv {\rm Arg}[(q/p)(\overline{\cal A}/{\cal A})]\approx {\rm Arg}(q/p)$ 
is ``universal,'' i.e., common to all $D^0$ decay modes. From 
Eq.~(\ref{eqn:acp_ind}) this implies that $a^{\rm indir}_{CP}$ 
is also universal. On the other hand, $a^{\rm dir}_{CP}$ depends 
on the final state. 

The relations between the observables and parameters are, to subleading 
order~\cite{Gersabeck},
\begin{eqnarray}
A^{}_\Gamma & \approx & -a^{\rm indir}_{CP} - a^{\rm dir}_{CP}\,y\cos\phi
\label{eqn:dcpv_ag} \\
A^{hh}_{CP} & \approx & 
a^{\rm dir}_{CP} - A^{}_\Gamma \frac{\langle t\rangle}{\tau}\,,
\label{eqn:dcpv_acp}
\end{eqnarray}
where $\langle t\rangle$ is the mean decay time for
$D^0\ra h^+h^-$ and $\tau$ is the $D^0$ lifetime. The
second term in Eq.~(\ref{eqn:dcpv_ag}) is the subleading
contribution -- compare to Eq.~(\ref{eqn:agamma}). It is
$O(10^{-4})$ or smaller and usually neglected; i.e.,
$A^{}_\Gamma\approx -a^{\rm indir}_{CP}$ and is considered 
universal. Inserting Eq.~(\ref{eqn:dcpv_ag}) into 
Eq.~(\ref{eqn:dcpv_acp}) gives 
\begin{eqnarray}
A^{hh}_{CP} & = & a^{\rm dir}_{CP} + a^{\rm indir}_{CP} \frac{\langle t\rangle}{\tau}
+ a^{\rm dir}_{CP} y\cos\phi \frac{\langle t\rangle}{\tau}\,.
\label{eqn:dcpv_new}
\end{eqnarray}

Experimentally, many systematic errors cancel when measuring
the difference $\Delta A^{}_{CP} \equiv A^{KK}_{CP} - A^{\pi\pi}_{CP}$.
Using Eq.~(\ref{eqn:dcpv_new}) to calculate this difference, one obtains
\begin{eqnarray}
\Delta A^{}_{CP} & = & 
 \Delta a^{\rm dir}_{CP} + a^{\rm indir}_{CP} \frac{\Delta \langle t\rangle}{\tau}
+ \left( a^{KK {\rm\ dir}}_{CP}\,\frac{\langle t\rangle_{KK}}{\tau}-
a^{\pi\pi {\rm\ dir}}_{CP}\,\frac{\langle t\rangle_{\pi\pi}}{\tau}\right) y\cos\phi \\
 & = & 
 \Delta a^{\rm dir}_{CP} \left( 1+ y\cos\phi \frac{\overline{\langle t\rangle}}{\tau}\right)
+ \left( a^{\rm indir}_{CP} + \overline{a^{\rm dir}_{CP}}\,y\cos\phi\right)
\frac{\Delta \langle t\rangle}{\tau} \\
 & \approx & 
 \Delta a^{\rm dir}_{CP} \left( 1+ y\cos\phi \frac{\overline{\langle t\rangle}}{\tau}\right)
+  a^{\rm indir}_{CP}\,\frac{\Delta \langle t\rangle}{\tau}\,,
\label{eqn:dcpv_dacp_final}
\end{eqnarray}
where 
$\Delta a^{\rm\ dir}_{CP} \equiv 
a^{KK {\rm\ dir}}_{CP} - a^{\pi\pi {\rm\ dir}}_{CP}$,
$\overline{a^{\rm\ dir}_{CP}} \equiv 
\left( a^{KK {\rm\ dir}}_{CP} + a^{\pi\pi {\rm\ dir}}_{CP}\right)/2$,
$\Delta \langle t\rangle \equiv 
\langle t\rangle_{KK} - \langle t\rangle_{\pi\pi}$, and 
$\overline{\langle t\rangle} \equiv 
\Bigl(\langle t\rangle_{KK} + \langle t\rangle_{\pi\pi}\Bigr)/2$.
Using Eq.~(\ref{eqn:dcpv_dacp_final}) 
and $A^{}_\Gamma = -a^{\rm indir}_{CP}$,
one can fit the measured values of $A^{}_\Gamma$ and 
$\Delta A^{}_{CP}$ for parameters $a^{\rm dir}_{CP}$ and
$a^{\rm indir}_{CP}$. A deviation from zero of either of
these parameters would indicate \cpv\ in $D$ decays. 
Such an observation would hint at new physics. To 
perform this fit requires knowledge of 
$\overline{\langle t\rangle}$, 
$\Delta \langle t\rangle$, and $y\cos\phi$.

\begin{figure}[htb]
\centering
\includegraphics[width=4.8in]{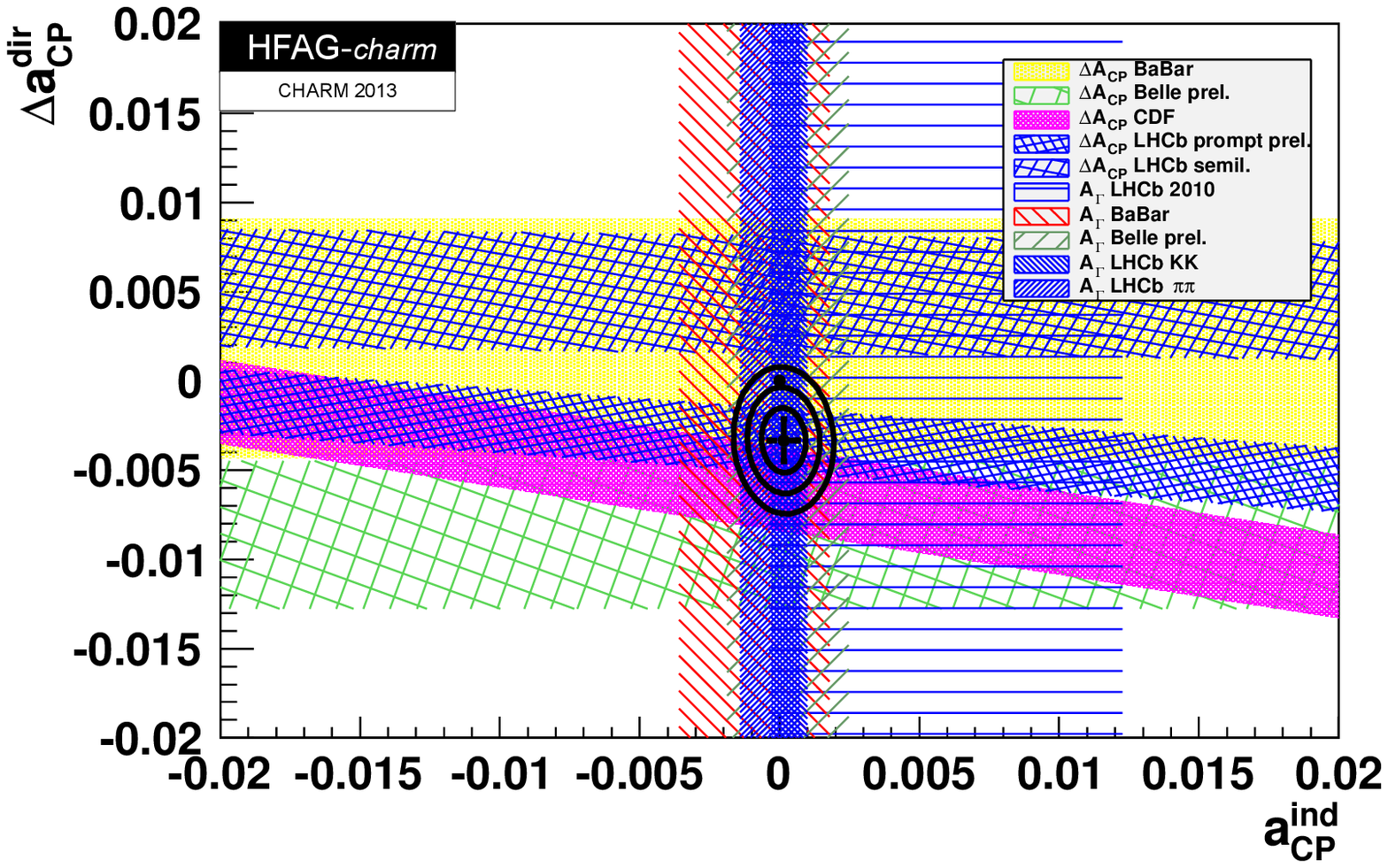}
\vskip-0.10in
\caption{Two-dimensional $\Delta a^{\rm dir}_{CP}$-$a^{\rm indir}_{CP}$ 
plane with constraints from individual experiments overlaid, from 
Ref.~\cite{hfag_dcpv}. The cross denotes the fitted central value, 
and the ellipses denote $1\sigma$, $2\sigma$, and $3\sigma$ 
confidence regions.}
\label{fig:Acp_hfag}
\end{figure}

The Heavy Flavor Averaging Group (HFAG)~\cite{hfag_web} performs 
this fit for all available data: Belle, BaBar, CDF, and
LHCb measurements. They use values of $\overline{\langle t\rangle}$
and $\Delta \langle t\rangle$ specific to each experiment, 
and $y\cos\phi$ is calculated using world average values~\cite{hfag_mixing}.
The resulting fit is shown in Fig.~\ref{fig:Acp_hfag}, 
which plots all relevant measurements in the two-dimensional
$\Delta a^{\rm dir}_{CP}$-$a^{\rm indir}_{CP}$ plane. The most 
likely values and $\pm 1\sigma$ errors are~\cite{hfag_dcpv}
\begin{eqnarray}
a^{\rm indir}_{CP} & = & (+0.015\pm 0.052)\% \\ 
\Delta a^{\rm dir}_{CP} & = & (-0.333\pm 0.120)\%\,.
\end{eqnarray}
Whereas $a^{\rm indir}_{CP}$ is consistent with zero,
$\Delta a^{\rm dir}_{CP}$ indicates direct \cpv. The 
$p$-value of the no-\cpv\ point 
$(a^{\rm indir}_{CP}, \Delta a^{\rm dir}_{CP})=(0,0)$ is~0.02.

\section{\boldmath Direct \cpv\ in $D^+\ra K^0_S\,K^+$}
\label{sec:won_ksk}

The decay $D^+\ra K^0_S\,K^+$ is self-tagging, there is no 
$D^{*\pm}\ra D\pi^\pm$ decay, and thus there is no correction 
for~$A^\pi_\varepsilon$. However, the final state $K^\pm$ 
introduces a correction ($A^{K^+}_\varepsilon$) due to possible 
differences in $K^+$ and $K^-$ reconstruction efficiencies. 
In addition, as the neutral $K^0$ or $\kbar$ is reconstructed 
via $K^0_S\ra\pi^+\pi^-$ decay, there is an asymmetry ($\acpKz$) 
due to the small difference in rates between $K^0\ra K^0_S$ and 
$\kbar\ra K^0_S$, or equivalently between $K^0\ra\pi^+\pi^-$ 
and $\kbar\ra\pi^+\pi^-$~\cite{GrossmanNir}.
Thus:
\begin{eqnarray*}
A^{K_S K^+}_{\rm recon} & = & 
\acpKzbKp + \acpKz + A^{}_{FB} + A^{K^+}_\varepsilon\,.
\end{eqnarray*}
Belle has measured $A^{K_S K^+}_{\rm recon}$ using 977~fb$^{-1}$ 
of data~\cite{belle_jhep02098} to determine $\acpKzbKp$.

\begin{figure}[htb]
\vskip0.10in
\centering
\includegraphics[width=2.75in]{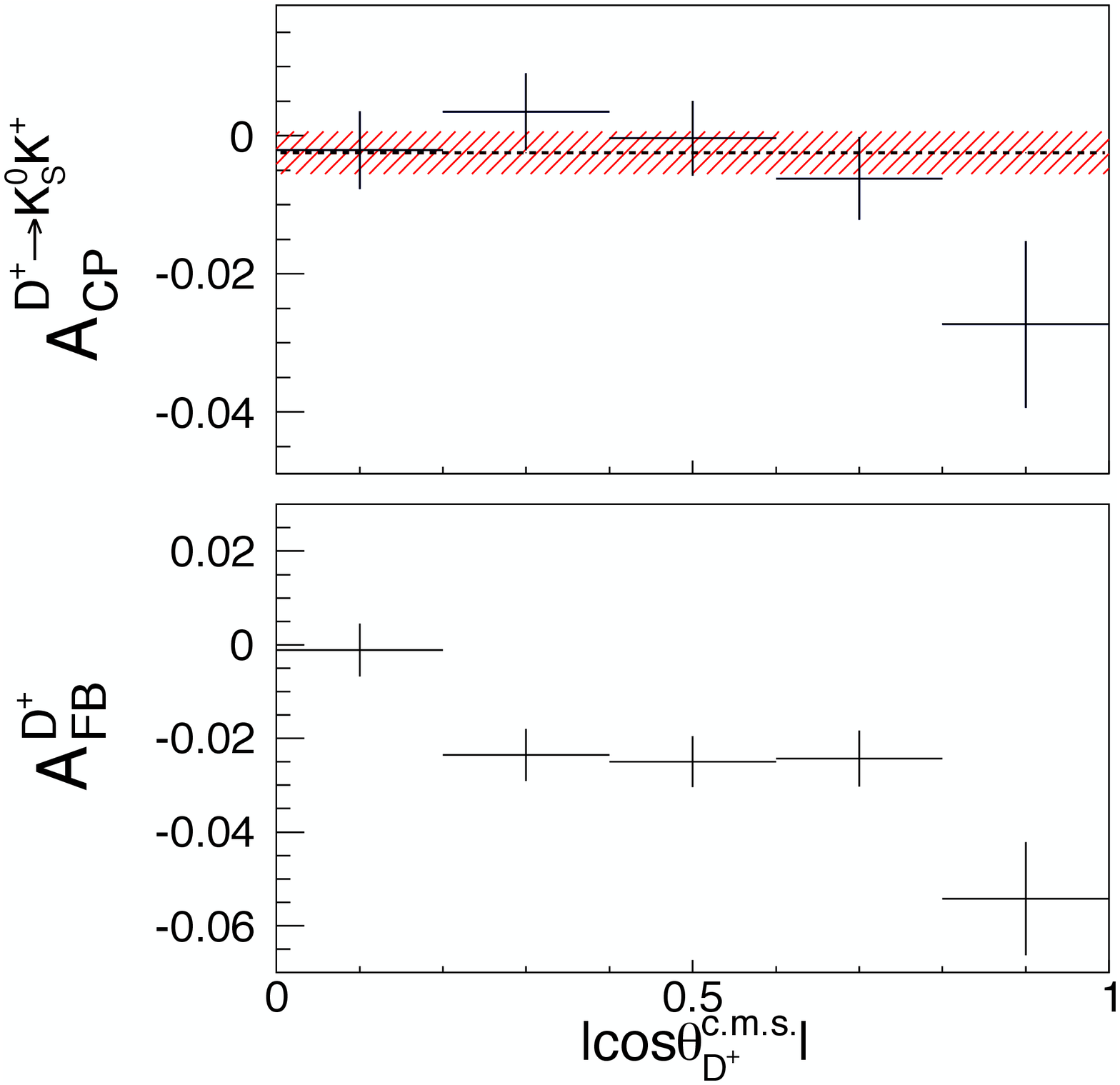}
\vskip-0.10in
\caption{$A^{K_S K^+}_{CP} = \acpKzbKp + \acpKz$ (top) and 
$A^{}_{FB}$ (bottom) measured from $D^\pm\ra K^0_S K^\pm$ 
decays~\cite{belle_jhep02098}.}
\label{fig:Acp_K0K}
\end{figure}

To determine $A^{K^+}_\varepsilon$, Belle measures the asymmetries for
untagged samples of $D^0\ra K^-\pi^+$ and $D_s^+\ra\phi \pi^+$ decays. 
These asymmetries have the following components:
\begin{eqnarray}
A^{K^-\pi^+}_{\rm recon} & = & A^{}_{FB} + 
            A^{\pi^+}_\varepsilon - A^{K^+}_\varepsilon \\ 
A^{\phi \pi^+}_{\rm recon} & = & A^{}_{FB} + A^{\pi^+}_\varepsilon\,.
\end{eqnarray}
To isolate $A^{K^+}_\varepsilon$, the weighting procedure 
performed for time-integrated $D^0\ra K^+K^-$ decays 
(see Section~\ref{sec:staric_ti}) is repeated here:
$D^0\ra K^-\pi^+$ decays are weighted by a factor
$(1-A^{\phi\pi^+}_{\rm recon})$, and $\dbar\ra K^+\pi^-$ decays 
are weighted by a factor $(1+A^{\phi\pi^+}_{\rm recon})$.
With this weighting the asymmetry $A^{K^-\pi^+}_{\rm recon}$ is 
calculated; the result equals $-A^{K^+}_\varepsilon$. This 
procedure is then repeated for the signal sample:
$D^+\ra K^0_S K^+$ decays are weighted by a factor
$(1-A^{K^+}_\varepsilon)$, and $D^-\ra K^0_S K^-$ decays 
are weighted by a factor $(1+A^{K^+}_\varepsilon)$.
The resulting $A^{K_S K^+}_{\rm recon}$ equals
$\acpKzbKp + \acpKz + A^{}_{FB}$.
Since the sum $\acpKzbKp + \acpKz$ is an even 
function of $\cos\theta^*$ ($\theta^*$ is the polar angle
with respect to the $e^+$ beam of the $D^+$ in the
$e^+e^-$ CM frame), and $A^{}_{FB}$ is an odd function, 
the two types of asymmetries are separated by taking
sums and differences as done in Eqs.~(\ref{eqn:acp}) 
and~(\ref{eqn:afb}). The results are plotted in 
Fig.~\ref{fig:Acp_K0K} in bins of $\cos\theta^*$; 
fitting these values to a constant yields 
\begin{eqnarray}
\acpKzbKp + \acpKz & = & (-0.25 \pm 0.28 \pm 0.14)\%\,.
\end{eqnarray}
To extract $\acpKzbKp$ from $\acpKzbKp + \acpKz$, one 
corrects for $\acpKz$ using the calculation of 
Ref.~\cite{GrossmanNir}. The result is 
\begin{eqnarray}
\acpKzbKp & = & (+0.08 \pm 0.28 \pm 0.14)\%\,,
\end{eqnarray}
which is consistent with zero.

\section{\boldmath Direct \cpv\ in $D^+\ra K^0_S\,\pi^+$}
\label{sec:won_ksp}

The decay $D^+\ra K^0_S\,\pi^+$ is similar to $D^+\ra K^0_S\,K^+$ 
in that it is also self-tagging and thus has no correction for
$D^{*\pm}\ra D\pi^\pm$. However, the final state $\pi^+$ 
introduces a correction $A^{\pi^+}_\varepsilon$ due to possible 
differences between $\pi^+$ and $\pi^-$ reconstruction efficiencies. 
The asymmetry $\acpKz$ is also present and must be corrected
for as done for $D^+\ra K^0_S\,K^+$ decays. Thus:
\begin{eqnarray}
A^{K_S\pi^+}_{\rm recon} & = & 
\acpKzbPp + \acpKz + A^{}_{FB} + A^{\pi^+}_\varepsilon\,.
\end{eqnarray}
Belle has measured $A^{K_S\pi^+}_{\rm recon}$ using 977~fb$^{-1}$ 
of data~\cite{belle_prl109} to determine $\acpKzbPp$.

\begin{figure}[htb]
\vskip0.10in
\centering
\includegraphics[width=2.75in]{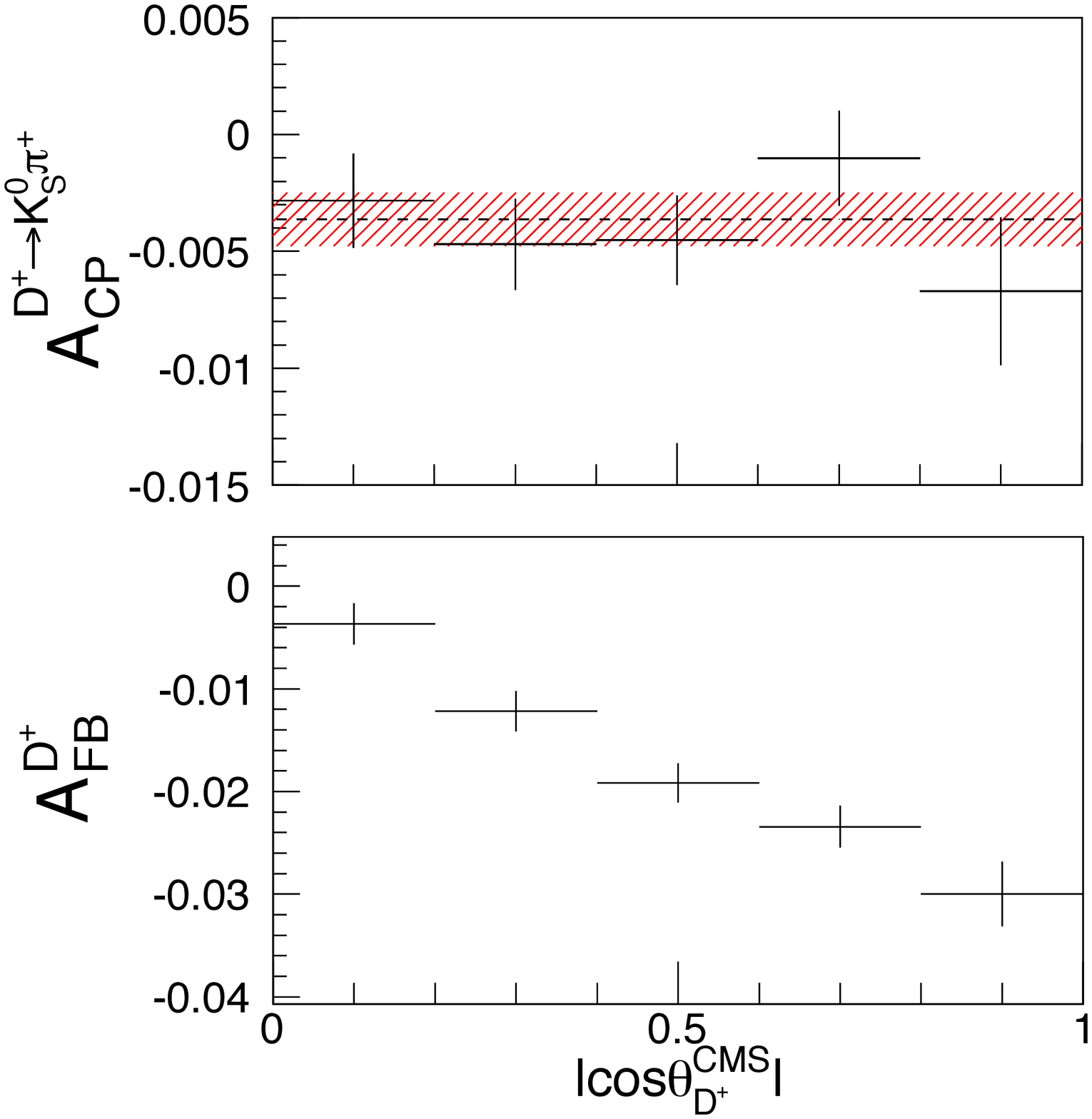}
\vskip-0.10in
\caption{$A^{K_S\,\pi^+}_{CP} = \acpKzbPp + \acpKz$ (top) and 
$A^{}_{FB}$ (bottom) measured from $D^\pm\ra K^0_S\,\pi^\pm$ 
decays~\cite{belle_prl109}.}
\label{fig:Acp_K0pi}
\end{figure}

To determine $A^{\pi^+}_\varepsilon$, Belle measures the 
asymmetries for untagged samples of three-body $D^+\ra K^-\pi^+\pi^+$ 
and $D^0\ra K^-\pi^+\pi^0$ decays.  These asymmetries have
the following components:
\begin{eqnarray}
A^{K^-\pi^+\pi^+}_{\rm recon} & = & A^{}_{FB} + 
          A^{K^-\pi^+}_\varepsilon + A^{\pi^+}_\varepsilon \\
A^{K^-\pi^+\pi^0}_{\rm recon} & = & A^{}_{FB} + A^{K^-\pi^+}_\varepsilon\,.
\end{eqnarray}
To isolate $A^{\pi^+}_\varepsilon$, the weighting procedure 
done for time-integrated $D^0\ra K^+K^-$ decays 
(Section~\ref{sec:staric_ti}) and $D^+\ra K^0_S K^+$ 
decays (Section~\ref{sec:won_ksk}) is repeated again:
$D^+\ra K^-\pi^+\pi^+$ decays are weighted by a factor
$(1-A^{K^-\pi^+\pi^0}_{\rm recon})$, and 
$D^-\ra K^+\pi^-\pi^-$ decays are weighted by a factor
$(1+A^{K^-\pi^+\pi^0}_{\rm recon})$. The resulting asymmetry
$A^{K^-\pi^+\pi^+}_{\rm recon}$ equals~$A^{\pi^+}_\varepsilon$. 
The weighting procedure is then repeated for the signal sample:
$D^+\ra K^0_S\,\pi^+$ decays are weighted by a factor
$(1-A^{\pi^+}_\varepsilon)$, and $D^-\ra K^0_S \pi^-$ decays 
are weighted by a factor $(1+A^{\pi^+}_\varepsilon)$.
The resulting $A^{K_S\pi^+}_{\rm recon}$ equals
$\acpKzbPp + \acpKz + A^{}_{FB}$.
Taking sums and differences in bins of $\cos\theta^*$
as done in Eqs.~(\ref{eqn:acp}) and (\ref{eqn:afb}) isolates 
$\acpKzbPp + \acpKz$. The results are plotted in 
Fig.~\ref{fig:Acp_K0pi}; fitting these values to 
a constant yields 
\begin{eqnarray}
\acpKzbPp + \acpKz & = & (-0.363\pm 0.094\pm 0.067)\%\,.
\end{eqnarray}
To extract $\acpKzbPp$, one applies a correction 
for $\acpKz$~\cite{GrossmanNir}. The result is
\begin{eqnarray}
\acpKzbPp & = & (-0.024\pm 0.094\pm 0.067)\%\,.
\end{eqnarray}
The statistics of this measurement are high enough to observe 
the asymmetry due to $\acpKz$ with a significance 
of~$3.2\sigma$. However, after correcting for $\acpKz$
the result for $\acpKzbPp$ is consistent with zero.

\section{\boldmath Other Searches for Direct \cpv}

In addition to the searches described above, there have been 
numerous other searches for direct \cpv\ in $D$ decays at Belle.
A complete listing of these searches and their results is given 
in Table~\ref{tab:direct_cpv_searches}. In all cases 
there is no evidence for direct \cpv.

\begin{table}[t]
\begin{center}
\begin{tabular}{l|ccc}
\hline
\textbf{Decay mode} & \textbf{\boldmath Data} & 
\textbf{\boldmath $A^{}_{CP}$ (\%)} & \textbf{Reference} \\
\hline
$D^0\ra\pi^+\pi^-$ & 977~fb$^{-1}$ & $(+0.55\pm 0.36\pm 0.09)$\% & \cite{belle_staric_acp} \\
$D^0\ra K^+ K^-$ & 977~fb$^{-1}$ & $(-0.32\pm 0.21\pm 0.09)$\% & \cite{belle_staric_acp} \\
$D^0\ra K^0_S\,\pi^0$ & 791~fb$^{-1}$ & $(-0.28\pm 0.19\pm 0.10)$\% & \cite{belle_prl106} \\
$D^0\ra K^0_S\,\eta$ & 791~fb$^{-1}$ & $(0.54\pm 0.51\pm 0.16)$\% & \cite{belle_prl106} \\
$D^0\ra K^0_S\,\eta'$ & 791~fb$^{-1}$ & $(0.98\pm 0.67\pm 0.14)$\% & \cite{belle_prl106} \\
$D^0\ra K^+\pi^-\pi^0$ & 281~fb$^{-1}$ & $(-0.6\pm 5.3)$\% & \cite{belle_prl95} \\
$D^0\ra K^+\pi^-\pi^+\pi^-$ & 281~fb$^{-1}$ & $(-1.8\pm 4.4)$\% & \cite{belle_prl95} \\
  &  &   &   \\
$D^+\ra K^0_S\,K^+$ & 977~fb$^{-1}$ & $(0.08\pm 0.28\pm 0.14)$\% & \cite{belle_jhep02098} \\
$D^+\ra K^0_S\,\pi^+$ & 977~fb$^{-1}$ & $(-0.024\pm 0.094\pm 0.067)$\% & \cite{belle_prl109} \\
$D^+\ra\phi\,\pi^+$ & 955~fb$^{-1}$ & $(0.51\pm 0.28\pm 0.05)$\% & \cite{belle_prl108} \\
$D^+\ra\pi^+\eta$ & 791~fb$^{-1}$ & $(1.74\pm 1.13\pm 0.19)$\% & \cite{belle_prl107} \\
$D^+\ra\pi^+\eta'$ & 791~fb$^{-1}$ & $(-0.12\pm 1.12\pm 0.17)$\% & \cite{belle_prl107} \\
  &  &   &   \\
$D^+_s\ra K^0_S\,\pi^+$ & 673~fb$^{-1}$ & $(5.45\pm 2.50\pm 0.33)$\% & \cite{belle_prl104} \\
$D^+_s\ra K^0_S\,K^+$ & 673~fb$^{-1}$ & $(0.12\pm 0.36\pm 0.22)$\% & \cite{belle_prl104} \\
\hline
\end{tabular}
\caption{Searches for direct \cpv\ in $D^0$, $D^+$, and $D^+_s$  decays at Belle.}
\label{tab:direct_cpv_searches}
\end{center}
\end{table}

\Acknowledgements
We are thankful to the organizers of CHARM 2013 for a fruitful and 
enjoyable workshop.


\begin{thebibliography}{99}
\bibitem{charge-conjugates}
Charge-conjugate modes are included unless stated or implied otherwise.
\end{thebibliography}
\end{document}